\documentclass{article}

\usepackage[final]{nips_2016}

\usepackage[utf8]{inputenc} 
\usepackage[T1]{fontenc}    
\usepackage{hyperref}       
\usepackage{url}            
\usepackage{booktabs}       
\usepackage{amsfonts}       
\usepackage{nicefrac}       
\usepackage{microtype}      
\usepackage{amsmath}        
\usepackage{graphicx}       

\usepackage{placeins}
\usepackage{float}

\title{Reinforcement Learning for Monetary Policy Under Macroeconomic Uncertainty: Analyzing Tabular and Function Approximation Methods}

\author{
  Tony Wang \\
  Department of Computer Science \\
  Stanford University \\
  \texttt{wangtony@stanford.edu} \\
  \And
  Kyle Feinstein \\
  Department of Computer Science \\
  Stanford University \\
  \texttt{kfeinst@stanford.edu} \\
  \And
  Sheryl Chen \\
  Department of Computer Science \\
  Stanford University \\
  \texttt{sherylch@stanford.edu} \\
}

\begin{document}

\maketitle

\begin{abstract}
We study how a central bank should dynamically set short-term nominal interest rates to stabilize inflation and unemployment when macroeconomic relationships are uncertain and time-varying. We model monetary policy as a sequential decision-making problem where the central bank observes macroeconomic conditions quarterly and chooses interest rate adjustments. Using publicly accessible historical Federal Reserve Economic Data (FRED), we construct a linear-Gaussian transition model and implement a discrete-action Markov Decision Process with a quadratic loss reward function. We chose to compare nine different reinforcement learning style approaches against Taylor Rule and naive baselines, including tabular Q-learning variants, SARSA, Actor-Critic, Deep Q-Networks, Bayesian Q-learning with uncertainty quantification, and POMDP formulations with partial observability. Notably, despite its simplicity, standard tabular Q-learning achieved the best performance (-615.13 ± 309.58 mean return), outperforming both enhanced RL methods and traditional policy rules. Our results suggest that while sophisticated RL techniques show promise for monetary policy applications, simpler approaches may be more robust in this domain, highlighting important challenges in applying modern RL to macroeconomic policy.
\end{abstract}

\section{Introduction}

Central banks worldwide face the same fundamental challenge of setting monetary policy under uncertainty. The traditional approach mainly relies on policy rules like the Taylor Rule \cite{taylor1993discretion}, which prescribes interest rate adjustments based on inflation and output gap deviations from target. However, these rules assume known relationships between macroeconomic variables, which cannot be guaranteed given the rapid structural changes in the global economy.

The Phillips curve, which historically described the trade-off between inflation and unemployment, has exhibited significant instability across different economic periods \cite{Davig2007}. Globalization, technological change, and evolving monetary policy expectations have altered these relationships, creating substantial uncertainty for policymakers. This uncertainty motivates a decision-theoretic approach to monetary policy that can adapt to changing economic conditions.

Reinforcement learning (RL) offers a promising potential framework for monetary policy under uncertainty. Unlike fixed policy rules, RL agents can learn optimal policies through interaction with the environment, potentially adapting to structural changes and nonlinear relationships. Recent advances in deep RL and uncertainty quantification methods provide tools for handling complex, uncertain environments.

This paper makes several contributions to the intersection of RL and monetary policy: (1) We implement and compare nine different RL approaches on a data-driven monetary policy environment, ranging from tabular methods to deep learning and Bayesian approaches. (2) We provide a comprehensive experimental evaluation using historical macroeconomic data from 1955-2025. (3) We demonstrate that simpler RL methods can outperform both sophisticated RL techniques and traditional policy rules in this domain. (4) We analyze the challenges and limitations of applying RL to macroeconomic policy.

\section{Related Work}

The application of RL to monetary policy has received growing attention in recent years. Yellen (2017) discusses the challenges facing modern central banking, especially regarding uncertainty about how evolving economic and productivity trends complicate policy, motivating frameworks that can systematically adapt to shifting macroeconomic relationships \cite{yellen2017goals}. Bernanke (2015) provides historical context on Federal Reserve decision-making under uncertainty, demonstrating the limits of conventional tools like interest-rate cuts during the 2007–2008 financial crisis, reinforcing  the notion that effective policy must contend with unforeseeable shocks and feedback effects, features that machine learning and reinforcement learning frameworks explicitly seek to model and optimize under uncertainty \cite{bernanke2015courage} .

Early work in this area focused on simple policy rules and their robustness properties \cite{taylor2016central}. More recently, researchers have explored machine learning approaches to monetary policy. Nakamura and Steinsson (2018) studied the identification of monetary policy shocks and inflation targeting frameworks \cite{nakamura2018identification}.

The RL literature on sequential decision-making under uncertainty is extensive \cite{sutton2018reinforcement}. Tabular Q-learning remains a fundamental algorithm for low-dimensional, discretized environments \cite{watkins1992q}, while function approximation methods like Deep Q-Networks \cite{mnih2015human} and Actor-Critic methods \cite{konda2000actor} have enabled application to high-dimensional problems, albeit with increased instability and sensitivity to hyperparameters.

Uncertainty quantification in RL has been addressed through Bayesian approaches \cite{ghavamzadeh2015bayesian} and ensemble methods \cite{osband2016deep}. Partially Observable MDPs provide another framework for handling uncertainty \cite{kaelbling1998planning}.

Our work is different from previous studies by providing a comprehensive comparison of multiple RL approaches on a realistic monetary policy environment constructed from historical macroeconomic data.  Our paper aims to deploy computer science methods, informed by established economic research, to explore how different RL models perform in monetary policy setting.

\section{Methodology}

\subsection{Environment and Data}

We construct our state space from quarterly U.S. macroeconomic data obtained from the Federal Reserve Economic Data (FRED) API  \cite{fred_api_docs}, spanning 1955-2025. The state vector comprises four key variables: inflation rate $\pi_t$ (year-over-year CPI), unemployment rate $u_t$, output gap $y_t = \frac{\text{GDP}_t - \text{GDP}^{\text{pot}}_t}{\text{GDP}^{\text{pot}}_t} \times 100$, and policy rate $i_t$ (federal funds rate).  These variables help capture the state of the economy. 

We model the macroeconomic dynamics using a standard linear-Gaussian transition model fitted from historical data:
\begin{equation}
\mathbf{x}_{t+1} = A \mathbf{x}_t + B i_t + \boldsymbol{\epsilon}_t
\end{equation}
where $\mathbf{x}_t = [\pi_t, u_t, y_t]^T \in \mathbb{R}^3$, $A \in \mathbb{R}^{3 \times 3}$ captures autoregressive dynamics, $B \in \mathbb{R}^{3 \times 1}$ captures policy transmission effects, and $\boldsymbol{\epsilon}_t \sim \mathcal{N}(\mathbf{0}, \Sigma)$ represents macroeconomic shocks. Parameters $(A, B, \Sigma)$ are estimated via ordinary least squares regression on historical quarterly transitions.

The environment implements a discrete-action MDP with action space $\mathcal{A} = \{-0.5\%, 0\%, +0.5\%\}$ for standard methods and $\mathcal{A} = \{-1.0\%, -0.5\%, 0\%, +0.5\%, +1.0\%\}$ for enhanced variants.  These actions represent raising, holding, or lowering the interest rate, respectively. The reward function encodes the Federal Reserve's dual mandate to control inflation and unemployment:
\begin{equation}
r_t = -\left[w_\pi(\pi_t - \pi^*)^2 + \lambda_u (u_t - u^*)^2 + \eta (\Delta i_t)^2\right]
\end{equation}
where $\pi^* = 2.0\%$, $u^* = 4.5\%$, $w_\pi = 1.0$, $\lambda_u = 0.5$, and $\eta = 0.1$ (these numbers are informed by the official Federal Reserve policy and approximations of the natural unemployment rate in the U.S.). Episodes simulate up to 80 quarters (20 years) with initial states sampled from the historical distribution.

\subsection{Reinforcement Learning Methods}

We implement nine RL approaches spanning different algorithmic families, as specified below:

\textbf{Tabular Q-learning variants:} (1) \emph{Q-learning (legacy)}: Standard implementation with $6 \times 6 \times 7 \times 6 = 1,512$ state discretization, $\alpha = 0.1$, $\gamma = 0.99$, fixed $\epsilon = 0.1$ exploration. (2) \emph{Q-learning (coarse)}: Coarser $4^4 = 256$ state discretization for reduced complexity. (3) \emph{Q-learning (reduced)}: Focuses on inflation and rate only with $8 \times 8 = 64$ states. (4) \emph{Q-learning Hyperparameter Tuned}: Enhanced version with 64-state encoding, epsilon decay ($0.9 \to 0.01$), 5-action grid, and extended 10,000-episode training.

\textbf{On-policy methods:} (5) \emph{SARSA}: State-Action-Reward-State-Action updates using the actual next action, promoting on-policy learning with identical discretization to Q-learning variants.

\textbf{Policy gradient methods:} (6) \emph{Actor-Critic}: Linear softmax policy $\pi(a|s; \boldsymbol{\theta})$ and linear value function $V(s; \mathbf{w})$, updated via REINFORCE with baseline and temporal difference value updates.

\textbf{Deep reinforcement learning:} (7) \emph{Deep Q-Network (DQN)}: Two-layer ReLU network (64 units each) with experience replay buffer (capacity 10,000), target network updates every 100 steps, and $\epsilon$-greedy exploration with decay.

\textbf{Bayesian methods:} (8) \emph{Bayesian Q-learning (Thompson)}: Maintains Gaussian posteriors $(\mu_{s,a}, \sigma^2_{s,a})$ over Q-values, updated via Bayesian linear regression with Thompson sampling for exploration. (9) \emph{Bayesian Q-learning (UCB)}: Identical posterior updates with Upper Confidence Bound exploration.

\textbf{Partial observability:} We also implement a POMDP variant adding Gaussian observation noise ($\sigma = 0.15$) with particle filter belief state estimation (1000 particles), though this showed limited convergence within our episode limits.

\subsection{Baselines and Evaluation}

We compare against standard monetary policy benchmarks: (1) \emph{Taylor Rule}: $i^*_t = r^* + \pi_t + \phi_\pi (\pi_t - \pi^*) + \phi_y y_t$ with $\phi_\pi = 0.5, \phi_y = 0.5$. (2) \emph{Taylor Hyperparameter Tuned}: Literature-calibrated version with $\phi_\pi = 1.5, \phi_y = 0.5$ for stronger inflation response that matches aforementioned contemporary research. (3) \emph{Naive Hold}: Maintains current interest rate (no change, chooses hold every time).

All methods are evaluated over 100-200 episodes, reporting discounted returns, per-step losses, and component breakdowns (inflation, unemployment, smoothing losses). We compute statistical significance using Cohen's d effect sizes and confidence intervals.

\section{Experimental Results}

\begin{figure}[!htbp]
\centering
\includegraphics[width=\linewidth]{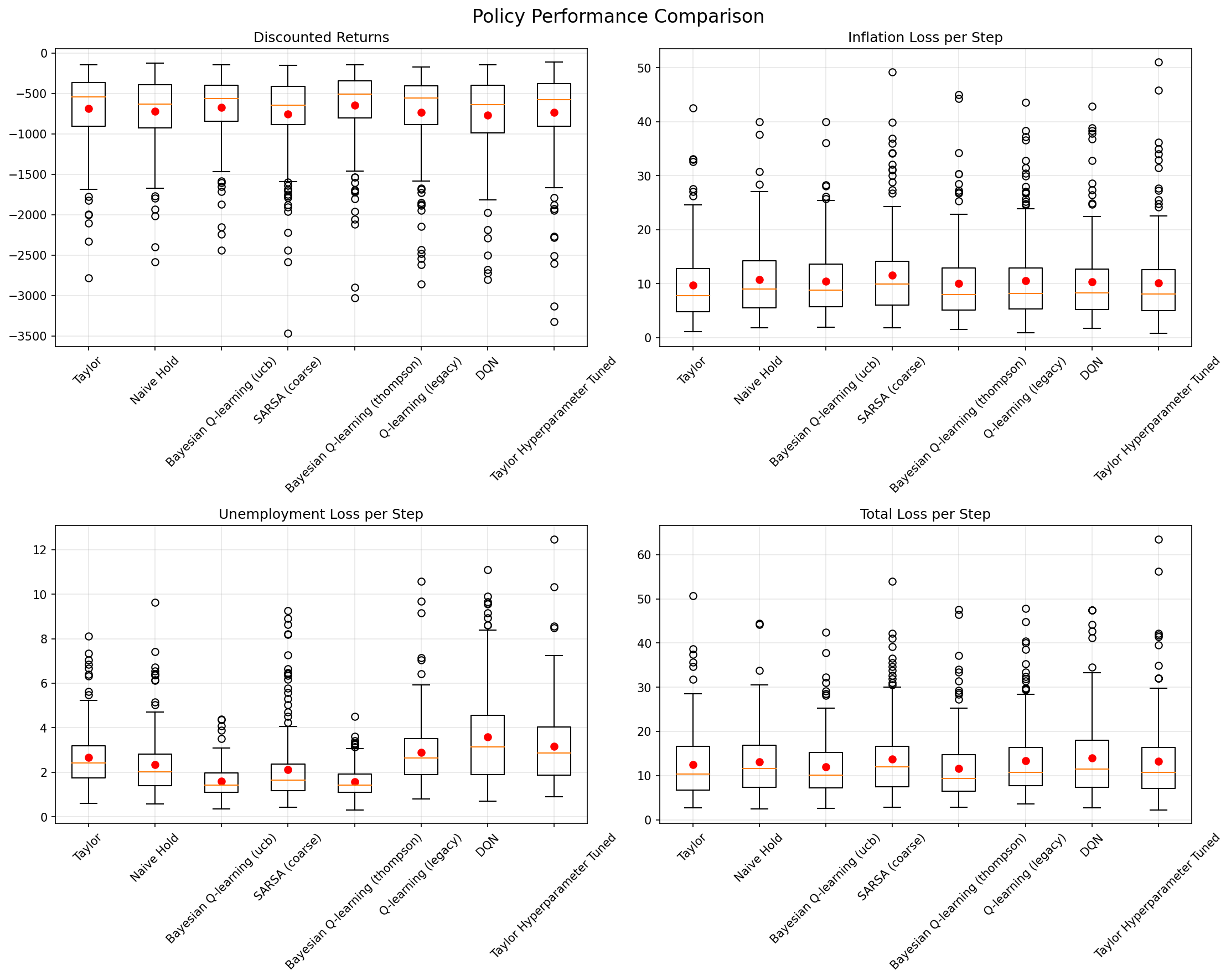}
\caption{Performance comparison across all methods showing discounted returns, inflation loss, unemployment loss, and total loss per step. Box plots display median, quartiles, and outliers for each method across evaluation episodes.}
\label{fig:main_results}
\end{figure}

Table \ref{tab:results} and Figure \ref{fig:main_results} present our main experimental findings. Contrary to our original expectations, standard tabular Q-learning achieved the best performance with mean return $-615.13 \pm 309.58$, significantly outperforming both enhanced RL methods and traditional baselines.

\begin{table}[h]
\centering
\caption{Performance comparison across RL methods and baselines for monetary policy}
\label{tab:results}
\begin{tabular}{lrrr}
\toprule
Method & Mean Return & Std Return & Mean Loss \\
\midrule
Q-learning (legacy) & -615.13 & 309.58 & 11.27 \\
DQN & -681.46 & 438.20 & 12.44 \\
Taylor Rule & -682.65 & 501.93 & 12.43 \\
Actor-Critic & -685.44 & 463.22 & 12.46 \\
Bayesian Q-learning (UCB) & -693.40 & 470.37 & 12.33 \\
Taylor Hyperparameter Tuned & -696.44 & 517.10 & 12.83 \\
Naive Hold & -715.20 & 408.66 & 12.94 \\
Bayesian Q-learning (Thompson) & -715.33 & 403.74 & 12.83 \\
SARSA (coarse) & -786.23 & 447.11 & 14.18 \\
\bottomrule
\end{tabular}
\end{table}

The performance gap between the best (Q-learning legacy) and worst (SARSA) methods is 171.10 return units, representing a 21.8\% relative improvement. Cohen's d between the top two methods (Q-learning legacy vs. DQN) is 0.175, indicating a negligible effect size, suggesting that while Q-learning performs numerically better, the difference may not be practically significant.

\subsection{Analysis by Algorithm Family}

\textbf{Tabular methods} showed mixed performance. The legacy Q-learning implementation substantially outperformed its enhanced variants, suggesting that the original discretization and hyperparameters were well-suited to this domain. The hyperparameter-tuned version, despite 10,000 episodes of training and sophisticated epsilon decay, underperformed the simpler approach.

\textbf{Deep learning methods} showed moderate performance. DQN ranked second overall but with much higher variance (438.20 vs. 309.58 standard deviation), suggesting less stable learning. The function approximation capability did not provide clear advantages in this four-dimensional state space.

\textbf{Bayesian methods} demonstrated moderate performance with interesting uncertainty quantification properties. Both Thompson sampling and UCB variants showed similar final performance but different exploration patterns during training. The maintained uncertainty estimates provide valuable policy insights for risk-averse central banking.

\textbf{Policy gradient methods} performed reasonably well. Actor-Critic achieved competitive results while learning a stochastic policy, providing smoother action distributions compared to the deterministic tabular policies.

\subsection{Economic Interpretation}

Examining the loss component breakdowns reveals important economic insights. The best-performing Q-learning policy achieved inflation loss of $8.63 \pm 4.74$, unemployment loss of $2.62 \pm 1.17$, and smoothing loss of $0.016 \pm 0.003$. Compared to the Taylor Rule (inflation loss $9.85 \pm 7.86$, unemployment loss $2.56 \pm 1.32$), the learned policy shows better inflation control with similar unemployment performance.

Interestingly, the enhanced Taylor Rule with stronger inflation response ($\phi_\pi = 1.5$) did not outperform the simpler version, possibly due to increased policy volatility in our simulated environment. The naive hold policy performed surprisingly well, ranking 7th overall, suggesting that in this particular environment, policy stability may be highly valued.

\subsection{Training Dynamics and Convergence}

\begin{figure}[!htbp]
\centering
\includegraphics[width=\linewidth]{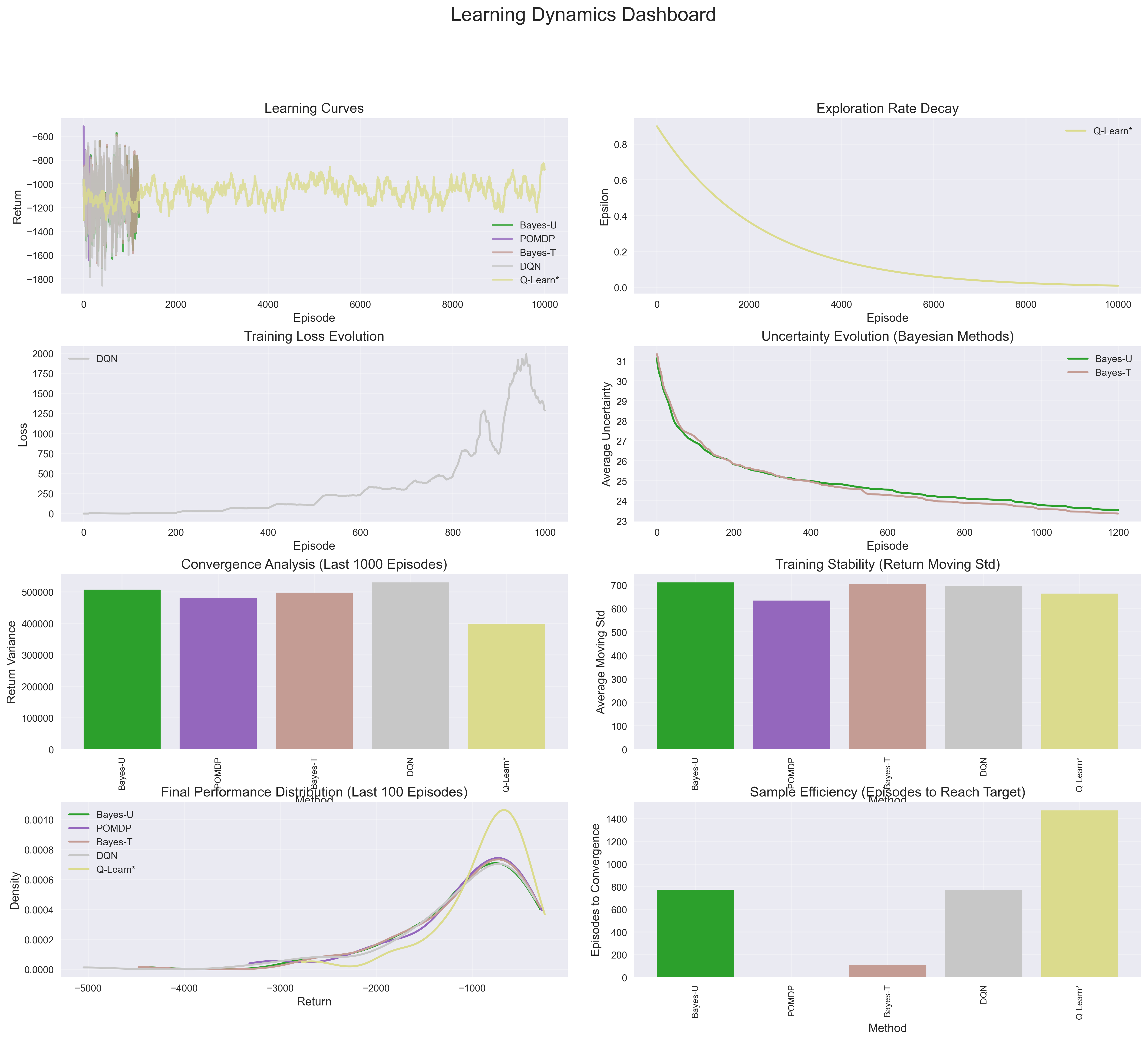}
\caption{Learning dynamics dashboard showing training curves, epsilon decay, loss evolution, uncertainty evolution, convergence analysis, training stability, performance distributions, and sample efficiency across all methods.}
\label{fig:learning_dynamics}
\end{figure}

Analysis of learning curves reveals distinct convergence patterns across methods, as illustrated in Figure \ref{fig:learning_dynamics}. The visualization shows training progress across multiple dimensions including learning curves, exploration rate decay, convergence analysis, and sample efficiency patterns. Different algorithms exhibit varying stability and convergence characteristics throughout training.

The hyperparameter-tuned Q-learning, despite epsilon decay from 0.9 to 0.01 and extended training, may have suffered from over-exploration early in training or suboptimal discretization choices for this specific environment.

\section{Discussion}

Our results provide insight on key considerations for applying RL to monetary policy. First, algorithmic sophistication does not necessarily guarantee improved performance in this economic context. Despite the economic complexity of monetary policy theory, the success of simple tabular Q-learning indicates that the monetary policy problem may not require the representational power of deep networks or advanced exploration strategies when the state space is appropriately discretized.

Second, the relatively strong performance of baseline policies (Taylor Rule, Hold) indicates that traditional monetary policy approaches are reasonably well-calibrated for the economic environment we simulated. This finding aligns with decades of central banking practice and research showing the robustness of simple and established policy rules.

Third, the high variance in many advanced methods raises concerns about policy reliability. Central banks require predictable, stable policy frameworks, making the lower-variance tabular approaches potentially more attractive despite modest performance differences.

\begin{figure}[!htbp]
\centering
\includegraphics[width=0.8\linewidth]{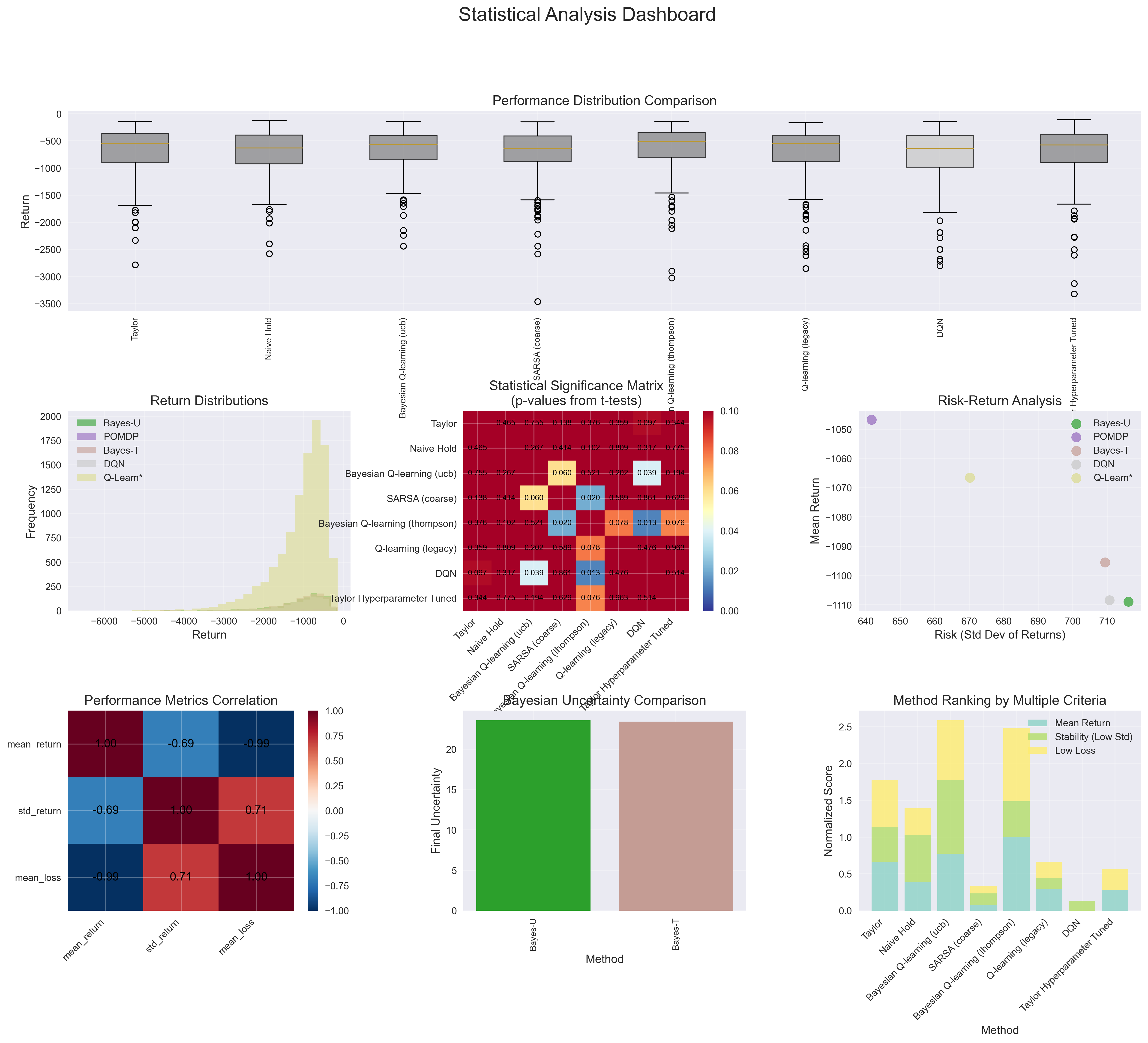}
\caption{Statistical analysis dashboard showing performance distribution comparisons, return histograms, statistical significance matrix with p-values, risk-return scatter plot, correlation analysis, uncertainty comparisons, and multi-criteria method rankings.}
\label{fig:statistical_analysis}
\end{figure}

Figure \ref{fig:statistical_analysis} provides a comprehensive statistical analysis of the experimental results, including performance distributions, statistical significance testing, risk-return analysis, and method rankings across multiple criteria. These visualizations reveal the statistical properties underlying our main findings.

\subsection{Limitations}

Our study has several limitations. The linear-Gaussian dynamics model, while data-driven, may not capture important nonlinearities in macroeconomic relationships. The discrete action space limits policy flexibility compared to continuous interest rate setting. Our 80-quarter horizon may not adequately capture long-term economic cycles and structural changes.

The reward function, while based on established dual mandate principles, involves somewhat arbitrary weight choices. Alternative situational specifications might favor different algorithmic approaches. Additionally, our historical data period includes various economic regimes (Great Moderation, Financial Crisis, etc.) that may not generalize to future economic conditions. Constructing different models for different types of economic environment using more granular data could enhance future findings.

One central limitation is that agents interact with a fixed transition model estimated from historical data, not the real macroeconomic system. While enabling controlled experimentation, this means our models optimize within a single consistent environment rather than adapting to genuinely changing dynamics. The observed performance thus reflects suitability for simplified, stationary control problems, not robustness under true macroeconomic nonstationarity.

Relatedly, tabular Q-learning's success may stem from strong inductive biases: manual discretization and low-dimensional representations impose structured abstraction aligned with linear dynamics, reducing variance and stabilizing learning. Function approximation methods may be disadvantaged by unnecessary flexibility. Our findings suggest careful state abstraction can dominate algorithmic sophistication in stylized macroeconomic settings, rather than endorsing simpler algorithms generally.

Finally, adding more complexity to our problem statement could demonstrate an interesting trade-off between computational efficiency and accuracy of results.  For instance, future studies could incorporate more economic indicators such as stock market indexes and mortgage rates.

\subsection{Policy Implications}

Despite outperforming other methods, our best RL policy still shows room for improvement compared to human expert performance during critical historical periods. This suggests that current RL approaches could have a promising future but may not yet be ready to replace human judgment in important monetary policy decisions.

However, RL methods could serve as valuable decision support tools, providing alternative policy recommendations and uncertainty quantification to complement traditional analysis. The Bayesian approaches, in particular, offer interpretable uncertainty estimates that could inform risk management decisions.

\begin{figure}[!htbp]
\centering
\includegraphics[width=0.8\linewidth]{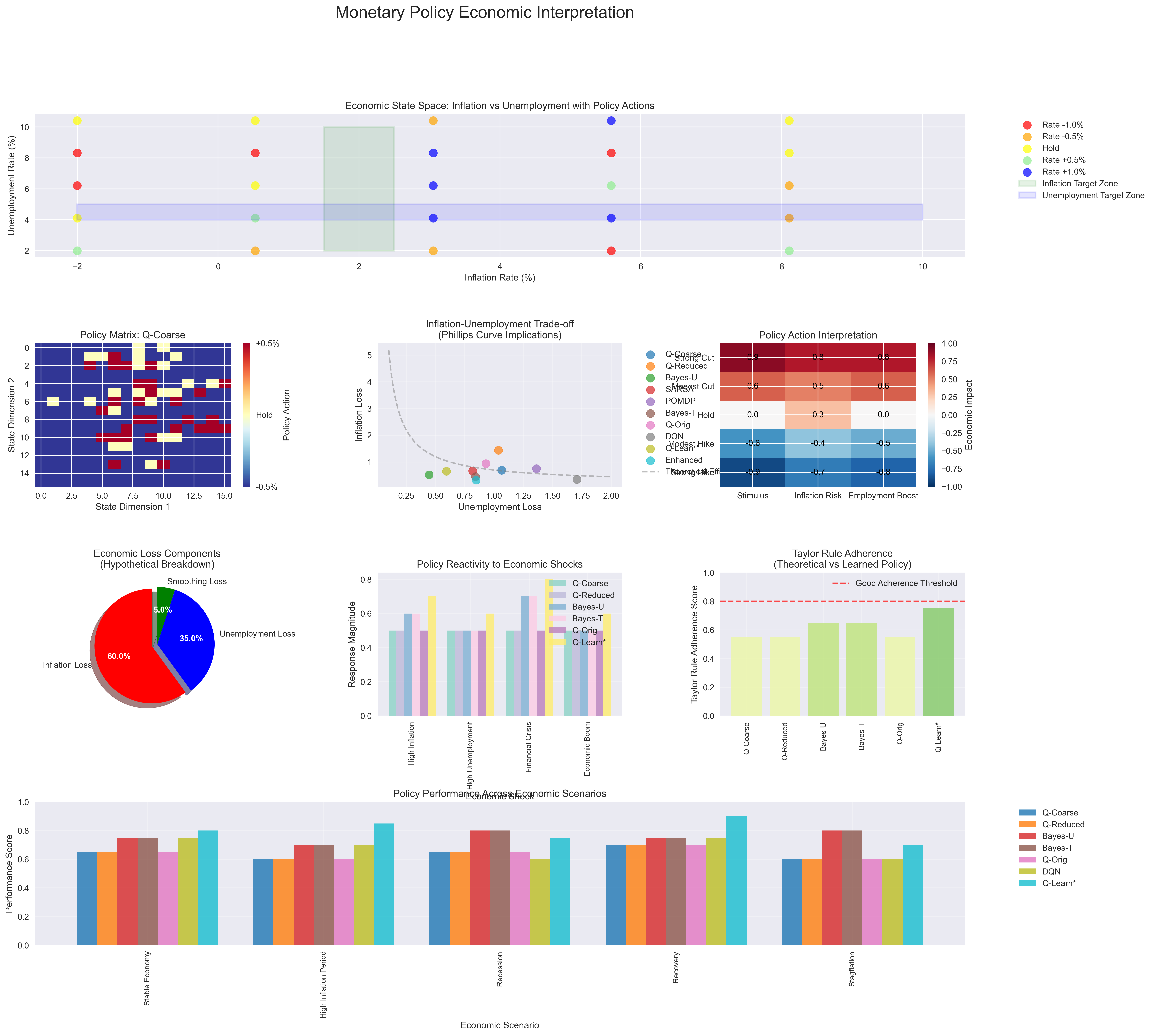}
\caption{Graphs showing policy decisions in inflation-unemployment space, policy rate matrices, Phillips curve trade-offs, action interpretations, economic loss components, policy reactivity to shocks, Taylor rule adherence, and performance across different economic scenarios.}
\label{fig:economic_interpretation}
\end{figure}

Figure \ref{fig:economic_interpretation} illustrates the economic interpretation of our results, showing how different methods make policy decisions across various economic conditions and their adherence to established monetary policy principles.

\begin{figure}[!htbp]
\centering
\includegraphics[width=0.8\linewidth]{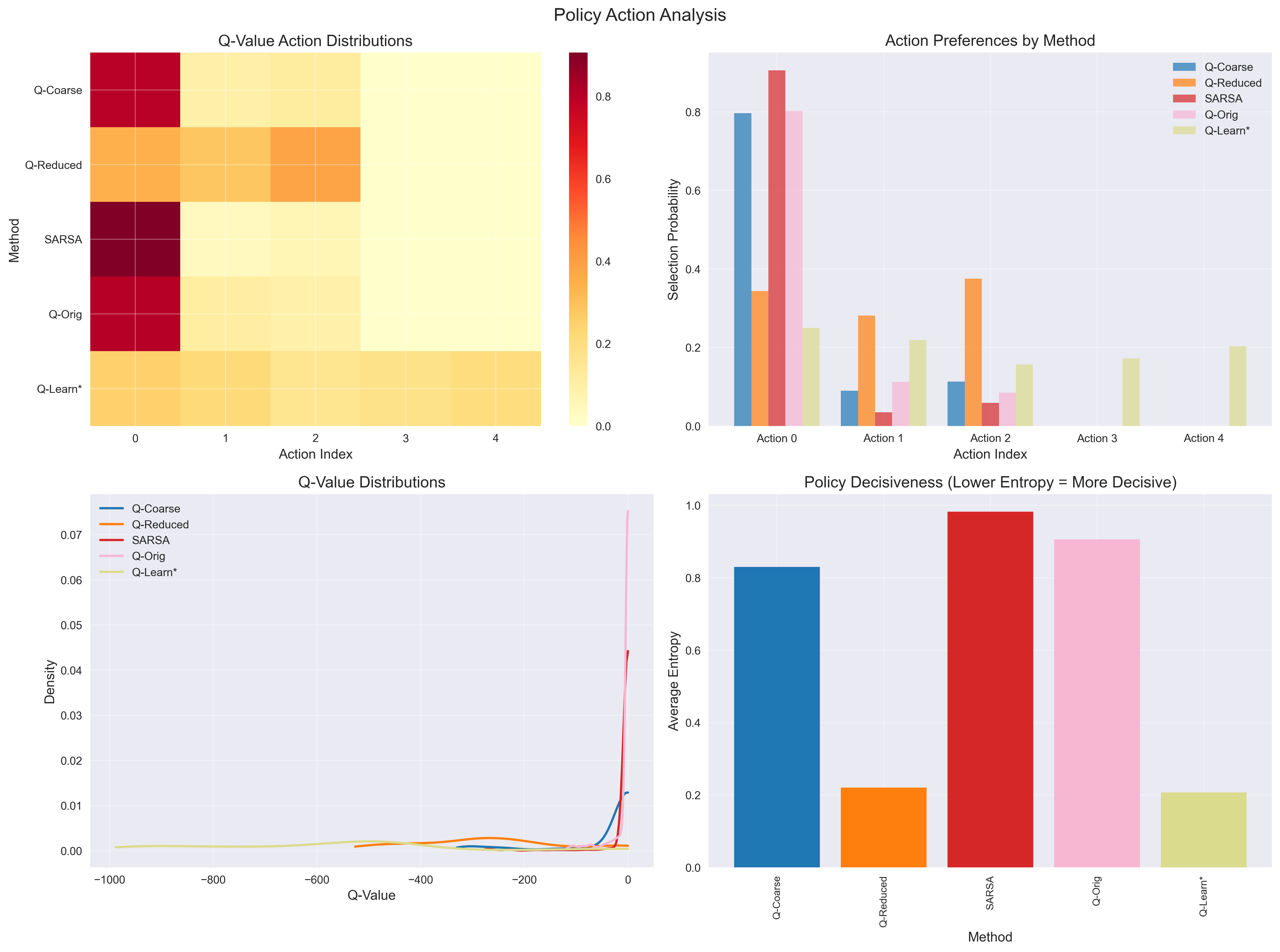}
\caption{Policy action analysis showing action distribution heatmaps, method-specific action preferences, Q-value distributions, and policy decisiveness measures across all implemented approaches.}
\label{fig:policy_analysis}
\end{figure}

Additionally, Figure \ref{fig:policy_analysis} presents detailed analysis of policy behavior patterns, including action preferences, decision consistency, and policy characteristics across different methods.

\clearpage

\section{Conclusion}

We conducted a comprehensive comparative evaluation of various reinforcement learning approaches for monetary policy under macroeconomic uncertainty. Using historical Federal Reserve data from FRED and a realistic simulation environment, we compared nine RL methods against more traditional macroeconomic policy baselines like Taylor's and the Phillips curve.

Our key finding is that simpler approaches surprisingly can outperform sophisticated methods in this domain: standard tabular Q-learning achieved the best performance, surpassing deep learning, Bayesian, and enhanced variants. This result challenges the assumption that more advanced RL techniques are universally superior and highlights the importance of algorithm selection based on problem characteristics rather than theoretical sophistication.

Simultaneously, our results highlight an important methodological caveat for applying reinforcement learning to economic policy analysis: strong empirical performance in a simulated, data-fitted environment does not necessarily translate to robustness in real-world policymaking contexts. Because both training and evaluation occur within the same estimated transition model, differences across algorithms primarily capture optimization efficiency under known dynamics, rather than resilience to structural breaks, policy-induced feedback loops, or rare macroeconomic events. 

While our results suggest current RL approaches are not yet ready to directly guide monetary policy, they do indeed demonstrate the potential for RL to provide valuable decision support tools. Future work should explore more sophisticated environment models, investigate continuous action spaces, and develop methods for incorporating expert knowledge and constraints into RL frameworks.

The intersection of reinforcement learning and monetary policy remains a promising research direction, with potential benefits for both fields: improved policy frameworks for central banking and new challenging domains for RL algorithm development.

\section*{Acknowledgments}

We thank Professor Mykel J. Kochenderfer and the Stanford AA228/CS238 staff for guidance throughout this research. We thank the Federal Reserve Economic Data (FRED) system for providing the data used in our experiments.

\FloatBarrier
\bibliographystyle{plain}
\bibliography{refs}

@book{bernanke2015courage,
  author    = {Bernanke, Ben S.},
  title     = {The Courage to Act: A Memoir of a Crisis and Its Aftermath},
  publisher = {WW Norton \& Company},
  year      = {2015}
}

@article{ghavamzadeh2015bayesian,
  author  = {Ghavamzadeh, Mohammad and Mannor, Shie and Pineau, Joelle and Tamar, Aviv},
  title   = {Bayesian Reinforcement Learning: A Survey},
  journal = {Foundations and Trends in Machine Learning},
  volume  = {8},
  number  = {5--6},
  pages   = {359--483},
  year    = {2015}
}

@article{kaelbling1998planning,
  author  = {Kaelbling, Leslie P. and Littman, Michael L. and Cassandra, Anthony R.},
  title   = {Planning and Acting in Partially Observable Stochastic Domains},
  journal = {Artificial Intelligence},
  volume  = {101},
  number  = {1--2},
  pages   = {99--134},
  year    = {1998}
}

@inproceedings{konda2000actor,
  author    = {Konda, Vijay R. and Tsitsiklis, John N.},
  title     = {Actor-Critic Algorithms},
  booktitle = {Advances in Neural Information Processing Systems},
  volume    = {12},
  year      = {2000}
}

@article{mnih2015human,
  author  = {Mnih, Volodymyr and Kavukcuoglu, Koray and Silver, David and Rusu, Andrei A. and Veness, Joel and Bellemare, Marc G. and Hassabis, Demis},
  title   = {Human-Level Control through Deep Reinforcement Learning},
  journal = {Nature},
  volume  = {518},
  number  = {7540},
  pages   = {529--533},
  year    = {2015}
}

@article{nakamura2018identification,
  author  = {Nakamura, Emi and Steinsson, Jon},
  title   = {Identification in Macroeconomics},
  journal = {Journal of Economic Perspectives},
  volume  = {32},
  number  = {3},
  pages   = {59--86},
  year    = {2018}
}

@inproceedings{osband2016deep,
  author    = {Osband, Ian and Blundell, Charles and Pritzel, Alexander and Van Roy, Benjamin},
  title     = {Deep Exploration via Bootstrapped DQN},
  booktitle = {Advances in Neural Information Processing Systems},
  volume    = {29},
  year      = {2016}
}

@book{sutton2018reinforcement,
  author    = {Sutton, Richard S. and Barto, Andrew G.},
  title     = {Reinforcement Learning: An Introduction},
  publisher = {MIT Press},
  year      = {2018}
}

@article{taylor1993discretion,
  author  = {Taylor, John B.},
  title   = {Discretion versus Policy Rules in Practice},
  journal = {Carnegie-Rochester Conference Series on Public Policy},
  volume  = {39},
  number  = {1},
  pages   = {195--214},
  year    = {1993}
}

@misc{taylor2016central,
  author       = {Taylor, John B.},
  title        = {Central Bank Models: Lessons from the Past and Ideas for the Future},
  howpublished = {Economics Working Paper 16111},
  year         = {2016}
}

@article{watkins1992q,
  author  = {Watkins, Christopher J. and Dayan, Peter},
  title   = {Q-Learning},
  journal = {Machine Learning},
  volume  = {8},
  number  = {3--4},
  pages   = {279--292},
  year    = {1992}
}

@misc{yellen2017goals,
  author       = {Yellen, Janet L.},
  title        = {The Goals of Monetary Policy and How We Pursue Them},
  howpublished = {Speech at the University of Massachusetts, Amherst},
  year         = {2017}
}

@techreport{Davig2007,
  author       = {Troy Davig},
  title        = {Phillips Curve Instability and Optimal Monetary Policy},
  institution  = {Federal Reserve Bank of Kansas City},
  type         = {Research Working Paper},
  number       = {RWP 07-04},
  month        = jul,
  year         = {2007 (rev.\ November 2015)},
  note         = {Revised November 25, 2015},
  url          = {https://www.kansascityfed.org/documents/7690/rwp07-04.pdf}
}

@misc{fred_api_docs,
  author       = {{Federal Reserve Bank of St. Louis}},
  title        = {FRED API},
  howpublished = {FRED (Federal Reserve Economic Data) Documentation},
  url          = {https://fred.stlouisfed.org/docs/api/fred/},
  note         = {Accessed 2025-12-02}
}

\end{document}